\documentclass[aps,preprint]{revtex4}%
\usepackage{amssymb}
\usepackage{amsfonts}
\usepackage{amsmath}
\usepackage{graphicx}%
\providecommand{\U}[1]{\protect\rule{.1in}{.1in}}

\begin{document}
\title[Hamiltonian relativity]{Kinematic relativity of quantum Hamiltonian}
\author{Gintautas P. Kamuntavi\v{c}ius}
\affiliation{Department of Physics, Vytautas Magnus University, Vileikos 8, Kaunas 44404, Lithuania}

\begin{abstract}
It is shown that the well-known relativistic correction of quantum Hamiltonian
that is present in textbooks appears after quantization of oversimplified
relativistic kinetic energy decomposition. Using the proper expression one
obtains the kinetic energy operator which relation with the ususal one is
consistent with corresponding relation of kinetic energies in special
relativity theory. The Schr\"{o}dinger equation with this operator gives new
interesting possibilities for quantum phenomena description.

\end{abstract}
\volumeyear{year}
\volumenumber{number}
\issuenumber{number}
\eid{identifier}
\date{2013.12.10}
\received[Received text]{date}

\revised[Revised text]{date}

\accepted[Accepted text]{date}

\published[Published text]{date}

\startpage{1}
\maketitle

Kinetic energy operator is the essential element of nonrelativistic quantum
Hamiltonian. Namely this operator defines the type of equation. By definition,
this operator appears from classical kinetic energy expression $T_{0}=$
$p_{0}^{2}/2m$, when instead of momentum $p_{0}=mv$\ corresponding quantum
operator%
\begin{equation}
\mathbf{p}_{0}=-i\hbar\nabla. \label{p0}%
\end{equation}
is applied. Relativistic corrections for kinetic energy operator are needed
for minimal modification of this Hamiltonian. The basic is the kinetic energy
definition in Special Relativity Theory (SRT):%
\begin{equation}
T=mc^{2}\sqrt{1+\left(  p/mc\right)  ^{2}}-mc^{2}. \label{RelT}%
\end{equation}
which expansion in terms of $p/mc$\ is%
\begin{equation}
T=\frac{p^{2}}{2m}\sum_{k=0}^{\infty}\frac{\left(  -1\right)  ^{k}}{\left(
k+1\right)  }\left(
\begin{array}
[c]{c}%
2k\\
k
\end{array}
\right)  \left(  \frac{p}{2mc}\right)  ^{2k}=\frac{p^{2}}{2m}-\frac{p^{4}%
}{8m^{3}c^{2}}+... \label{TsklK}%
\end{equation}
(here $\left(
\begin{array}
[c]{c}%
2k\\
k
\end{array}
\right)  $\ denotes the binomial coefficient). The first member of this
expansion is defined as nonrelativistic kinetic energy, and the second
provides the mentioned correction. Namely this definition is present in
quantum mechanics textbooks, beginning with classical, such as \cite{PAMD}%
,\cite{Mesiah},\cite{Schiff} and finishing with the modern ones \cite{Cohen}%
,\cite{Manou} or \cite{Henley}. The same expansion also appears in Dirac
equation transformation to the Schr\"{o}dinger form, hence it is widely
accepted in journals publications, considering the relativistic corrections of
different quantum Hamiltonians.

However, this conclusion follows from definition that at SRT kinetic energy
expression $\left(  \ref{RelT}\right)  $ quantization the operator
$-i\hbar\nabla$ corresponds to the momentum $p$\ so that the first term of
expansion $\left(  \ref{TsklK}\right)  $ equals the kinetic energy operator,
present in Schr\"{o}dinger equation. This means that instead of the
relativistic kinetic energy $\left(  \ref{RelT}\right)  $ simplified form%
\begin{equation}
mc^{2}\sqrt{1+\left(  p_{0}/mc\right)  ^{2}}-mc^{2}%
\end{equation}
is applied. As a result, the second term of expansion $\left(  \ref{TsklK}%
\right)  $ appears with negative sign, defining that the corrected kinetic
energy is smaller than the nonrelativistic one. Obviously, this result is far
from reality because the corrected value has to be larger than the
nonrelativistic ($T>T_{0}$ at all momentum values, see Eq.$\left(
\ref{TT0}\right)  $ ). The well known explanation of this improper conclusion
sounds like the one given in \cite{Cohen}, where the first term of expansion
is called as "The nonrelativistic kinetic energy", and the second as "The
first energy correction, due to the relativistic variation of the mass with
the velocity". The result obtained following the widely accepted
recommendations is problematic also from point of view of relativistic
character of momentum and corresponding operator. Nonrelativistic momentum as
function of particle velocity is linear at all values of argument. Behavior of
relativistic momentum in this situation is completely another - it takes
infinite value at $v=c$\ and is undefined for larger velocities. Corresponding
quantum operator has to have the analogous dependence, however the standard
well-known operator's $-i\hbar\nabla$ eigenvalues spectrum is without any
signs of relativity.

As follows from the definition, the problem is that the momentum present in
$\left(  \ref{RelT}\right)  $\ is the relativistic momentum $p=\gamma p_{0}%
,$\ where%
\begin{equation}
\gamma=1/\sqrt{1-\left(  p_{0}/mc\right)  ^{2}}.\ \label{Gamap0}%
\end{equation}
SRT kinetic energy expansion, which first term is namely nonrelativistic
kinetic energy, can be obtained applying the given above $\gamma$ and
definition, equivalent to $\left(  \ref{RelT}\right)  $ but rewritten as
\begin{equation}
T=mc^{2}\left(  \gamma-1\right)  .
\end{equation}
In this case one can easily obtain the following result:%
\begin{equation}
T=mc^{2}\sum_{k=1}^{\infty}\left(
\begin{array}
[c]{c}%
2k\\
k
\end{array}
\right)  \left(  \frac{p_{0}}{2mc}\right)  ^{2k}=\frac{p_{0}^{2}}{2m}%
+\frac{3p_{0}^{4}}{8m^{3}c^{2}}+.... \label{TT0}%
\end{equation}

Now, the first member of this expansion exactly corresponds the quantum
nonrelativistic kinetic energy operator%
\begin{equation}
\mathbf{T}_{0}=-\frac{\hbar^{2}}{2m}\nabla^{2}, \label{Tnulis}%
\end{equation}
while the second%
\begin{equation}
\frac{3}{2}\frac{\mathbf{T}_{0}^{2}}{mc^{2}}%
\end{equation}
can be applied for relativistic corrections of Schr\"{o}dinger equation
eigenvalues. Mean value of this operator, equal $3t_{0}^{2}/2mc^{2}$, at
$t_{0}\ll mc^{2}/2$ is small in comparison with nonrelativistic kinetic energy
$t_{0}$, obtained after Schr\"{o}dinger equation solution. However, this
correction can be remarkable in situation, when particle is loosely bound in
deep potential well. In this case the kinetic and also the potential energies
are large, but their difference, equal binding energy of state under
consideration, is small enough. Namely such problems are characteristic for
strong interaction potentials of microscopic nuclear theory.

However, this correction does not help taking into account the mentioned
relativistic momentum and kinetic energy $\left(  \ref{RelT}\right)  $
behavior at particle velocity, approaching $c.$

This feature of SRT can be taken into account after complete quantization of
the first term of expansion $\left(  \ref{TsklK}\right)  ,$ equal $T_{1}%
=p^{2}/2m$.\ Corrected expression for kinetic energy equals
\begin{equation}
T_{1}=\gamma^{2}\frac{p_{0}^{2}}{2m}=\frac{p_{0}^{2}/2m}{1-\left(
p_{0}/mc\right)  ^{2}}\equiv\frac{T_{0}}{1-2T_{0}/mc^{2}},
\end{equation}
and corresponding quantum mechanical operator is
\begin{equation}
\mathbf{T}_{1}=\left(  1-2\mathbf{T}_{0}/mc^{2}\right)  ^{-1}\mathbf{T}_{0}.
\label{KorH}%
\end{equation}

The eigenfunctions of this operator coincide with corresponding eigenfunctions
of operator $\mathbf{T}_{0},$ because the commutator of these operators equals
zero. However, every eigenfunction corresponds with different eigenvalues of
$\mathbf{T}_{0}$\ and $\mathbf{T}_{1}$. At nonrelativistic kinetic energy,
equal $t_{0},$ the corrected eigenvalue of operator $\mathbf{T}_{1}$ equals%
\begin{equation}
t_{1}=t_{0}/\left(  1-2t_{0}/mc^{2}\right)  .
\end{equation}
Due to positive definiteness of operator $\mathbf{T}_{1}$\ its eigenvalues
have to be positive, hence a condition for nonrelativistic eigenvalue
$t_{0}<mc^{2}/2$ appears.

Let us investigate the spectrum of this operator for particle in infinitely
deep spherically symmetric well. This is the problem of spherical cavity with
completely impenetrable walls. The potential inside of well equals zero and
outside of well equals infinity. Schr\"{o}dinger equation for this problem is
well known:%
\begin{equation}
\mathbf{T}_{0}\varphi_{nl\mu}\left(  \mathbf{r}\right)  =e_{nl}\varphi_{nl\mu
}\left(  \mathbf{r}\right)  . \label{NRT}%
\end{equation}
The boundary condition%
\begin{equation}
\varphi_{nl\mu}\left(  \left\vert \mathbf{r}\right\vert =R\right)  \equiv0,
\end{equation}
where the radius of well equals $R,$\ defines infinite set of this equation
solutions. Here $l\mu$\ denotes angular momentum and projection quantum
numbers, $n=1,2,...$\ is number of bound state with given $l$.

For corrected operator with the same boundary condition the corresponding
equation is%
\begin{equation}
\left(  1-2\mathbf{T}_{0}/mc^{2}\right)  ^{-1}\mathbf{T}_{0}\varphi_{nl\mu
}\left(  \mathbf{r}\right)  =E_{nl}\varphi_{nl\mu}\left(  \mathbf{r}\right)  .
\label{RRT}%
\end{equation}
The eigenfunctions of both equations - $\left(  \ref{NRT}\right)  $ and
$\left(  \ref{RRT}\right)  $ again coincide, the eigenvalues are different:%
\begin{equation}
E_{nl}=\frac{e_{nl}}{1-2e_{nl}/mc^{2}}.
\end{equation}
This relation clearly demonstrates the considered relativistic character of
the corrected problem. As mentioned, kinetic energy operator is positively
defined, hence in corrected operator spectrum only some of nonrelativistic
Hamiltonian eigenvalues\ not exceeding mentioned maximal possible kinetic
energy value, i.e. $e_{nl}<mc^{2}/2,$ are present. As a conclusion it follows
that in infinite well finite number of solutions exists.\ Moreover, when
radius of well stays smaller than some critical value, following from
condition $\min e_{nl}\equiv e_{10}=mc^{2}/2,$\ spectrum of this operator is
empty. For this radius definition one has to obtain minimal eigenvalue\ of
corresponding well. Obviously, it will have angular momentum, that equal zero.
Hence,%
\begin{equation}
e_{10}=\frac{mc^{2}}{2}\left(  \frac{\pi\hbar c}{Rmc^{2}}\right)  ^{2}.
\end{equation}
Mentioned condition gives the following value of radius:%
\begin{equation}
R=\frac{\pi\hbar c}{mc^{2}}.
\end{equation}
Applying $\hbar c=197.327$ $MeVfm$ for proton $\left(  mc^{2}=938.272\text{
}MeV\right)  $ the radius of empty well equals $R_{p}=0.660\,$\ $fm$. For
electron $\left(  mc^{2}=0.511\text{ }MeV\right)  $ this radius is
$R_{e}=1.212$ $pm$.

If the potential $v\left(  \mathbf{r}\right)  $ does not equal identical zero,
the Schr\"{o}dinger equation is%
\begin{equation}
\left(  1-2\mathbf{T}_{0}/mc^{2}\right)  ^{-1}\mathbf{T}_{0}\psi\left(
\mathbf{r}\right)  =\left(  E-v\left(  \mathbf{r}\right)  \right)  \psi\left(
\mathbf{r}\right)  .
\end{equation}
Due to mentioned condition for operator $\left(  1-2\mathbf{T}_{0}%
/mc^{2}\right)  $\ eigenvalues and obvious condition $\left(  1-2\mathbf{T}%
_{0}/mc^{2}\right)  \psi\left(  \mathbf{r}\right)  \neq0,$ this equation can
be present in a following form:%
\begin{equation}
\mathbf{T}_{0}\psi\left(  \mathbf{r}\right)  =\left(  1-2\mathbf{T}_{0}%
/mc^{2}\right)  \left(  E-v\left(  \mathbf{r}\right)  \right)  \psi\left(
\mathbf{r}\right)  ,
\end{equation}
or%
\begin{equation}
\left[  1-2v\left(  \mathbf{r}\right)  +2E\right]  \mathbf{T}_{0}\psi\left(
\mathbf{r}\right)  +mc^{2}\left[  v\left(  \mathbf{r}\right)  -E\right]
\psi\left(  \mathbf{r}\right)  -2\left[  \mathbf{T}_{0},v\left(
\mathbf{r}\right)  \right]  \psi\left(  \mathbf{r}\right)  =0 \label{Lygt}%
\end{equation}
with aforementioned condition for corrected kinetic energy value%
\begin{equation}
\int d\mathbf{r\psi}^{+}\mathbf{\left(  \mathbf{r}\right)  T}_{0}\psi\left(
\mathbf{r}\right)  <mc^{2}/2.
\end{equation}
The $\left[  \mathbf{T}_{0},v\left(  \mathbf{r}\right)  \right]  $ denotes
commutator of nonrelativistic kinetic energy $\left(  \ref{Tnulis}\right)  $
and potential operators.

The equation obtained $\left(  \ref{Lygt}\right)  $ is not trivial, its
solutions, i.e. spectrum and eigenfunctions for different potentials, can give
interesting information about relativistic effects in Schr\"{o}dinger
formalism at minimal relativity taken into account.

In conclusion one has to pay attention to the fact that factor $\gamma
^{2}=\left(  1-2\mathbf{T}_{0}/mc^{2}\right)  ^{-1}$\ appears in front of
entire relativistic kinetic energy expression $\left(  \ref{TsklK}\right)
,$\ hence the obtained results can be more fundamental than simple "kinematic
relativity" approach.\

\end{document}